\documentclass[prd,showpacs,amsmath,amssymb]{revtex4-1}
\usepackage{amssymb}
\usepackage{amsmath}
\usepackage{graphicx}
\usepackage{dcolumn}
\usepackage{bm}
\usepackage{epsfig}
\usepackage{amsfonts}%
\providecommand{\U}[1]{\protect\rule{.1in}{.1in}}
\def\br{\begin{eqnarray}}
\def\er{\end{eqnarray}}
\def\be{\begin{equation}}
\def\ee{\end{equation}}

\def\({\left(}
\def\){\right)}

\def\<{\left\langle}
\def\>{\right\rangle}

\begin{document}
\title{Limit on the pion distribution amplitude}
\author{E. G. S. Luna$^{1}$ and A. A. Natale$^{2,3}$}
\affiliation{$^{1}$Instituto de F\'{\i}sica, Universidade Federal do Rio Grande do Sul, Caixa
Postal 15051, 91501-970, Porto Alegre, RS, Brazil }
\affiliation{$^{2}$Centro de Ci\^encias Naturais e Humanas, Universidade Federal do ABC,
09210-170, Santo Andr\'e - SP, Brazil}
\affiliation{$^{3}$Instituto de F\'{\i}sica Te\'orica, UNESP, Rua Dr. Bento T.
Ferraz, 271, Bloco II, 01140-070, S\~ao Paulo - SP, Brazil }

\begin{abstract}
The pion distribution amplitude (DA) can be related to the fundamental QCD
Green's functions as a function of the quark self-energy and the quark-pion
vertex, which in turn are associated with the pion wave function through the
Bethe-Salpeter equation. Considering the extreme hard asymptotic behavior in
momentum space allowed for a pseudoscalar wave function, which is limited by
its normalization condition, we compute the pion DA and its second moment.
From the resulting amplitude, representing the field theoretical upper limit
on the DA behavior, we calculate the photon-pion transition form factor
$F_{\pi\gamma\gamma^{\ast}}(Q^{2})$. The resulting upper limit on the pion transition
form factor is compared with existing data published by CLEO, BaBar and Belle collaborations.

\end{abstract}
\maketitle


A few years ago new data were published \cite{:2009mc,Uehara:2012ag} for the
$\gamma^{\ast}\gamma\rightarrow\pi^{0}$ process, where one of the photons is
far off mass shell (large $Q^{2}$) and the other one is near mass shell
($Q^{2}\approx0$). These measurements of the photon-pion transition form
factor $F_{\pi\gamma\gamma^{\ast}}(Q^{2})$, taken in single-tagged two-photon
$e^{+}e^{-}\rightarrow e^{+}e^{-}\pi^{0}$ reaction, was performed in a wide
range of momentum transfer squared ($4-40$ GeV$^{2}$). At sufficiently high
$Q^{2}$ it is expected that the standard factorization approach can be applied
\cite{Chernyak:1977as,Lepage:1979zb,Efremov:1979qk} (for a review, see
\cite{bl}). The amplitude for this process at high virtuality has the form%

\begin{align}
F_{\pi\gamma\gamma^{*}}(Q^{2}) = \frac{2f_{\pi}}{3}\int_{0}^{1}dx \,
\varphi_{\pi}(x)T^{H}_{\gamma\pi}(x,Q^{2}). \label{eqa}%
\end{align}
This equation is obtained assuming factorization of the pion distribution
amplitude (DA) $\varphi_{\pi}(x)$ and the hard scattering amplitude
$T^{H}_{\gamma\pi}(x,Q^{2})$ given by \cite{brodsky,brodsky2}%

\begin{equation}
T^{H}_{\gamma\pi}(x,Q^{2}) = \frac{1}{(1-x)Q^{2}} [ 1 + \mathcal{O}(\alpha
_{s})]. \label{eqb}%
\end{equation}

However the BaBar Collaboration data \cite{:2009mc} seems to be in
contradiction with this approach since, in accordance with perturbative QCD
(pQCD), $Q^{2}F_{\pi\gamma\gamma^{\ast}}(Q^{2}\rightarrow\infty)$ should be
limited to the value $2f_{\pi}\approx0.185$ GeV \cite{brodsky2}, hereafter
called BL limit. At the same time, the Belle Collaboration data
\cite{Uehara:2012ag} presented in the same range of transferred momenta show
that the pion transition form factor may not increase as fast as shown by the
BaBar results.

As a consequence of these experiments there were many theoretical papers
speculating why the data should (or not) obey the BL limit
\cite{Radyushkin:2009zg,Polyakov:2009je,Dorokhov:2010zzb,Dorokhov:2011dp,Mikhailov:2009kf,Brodsky:2011yv,Arriola:2010aq,Noguera:2010fe,Agaev:2010aq,Kroll:2010bf,Klopot:2012hd}%
. The first attempt to explain the BaBar
result can be found in Ref. \cite{dorokha001}. Among these there were proposals claiming that the pion distribution
amplitude should be modified
\cite{Radyushkin:2009zg,Polyakov:2009je,Dorokhov:2010zzb,Noguera:2010fe}%
, leading to a broader or flatter distribution in the place of the asymptotic
form $\varphi_{\pi}^{as}(x)=6x(1-x)$ \cite{efr}. A flat DA would be consistent
with the BaBar data, although a field theoretical support for such possibility
is still missing. Some papers claim that other transition form factors of
heavier mesons are compatible among themselves and with the saturation
required by factorization theorems obtained from pQCD
\cite{Mikhailov:2009kf,ste}. However, for heavier mesons than the pion the DA
may be more peaked away from the end points \cite{Dorokhov:2011dp}. A common statement in
all papers is the need for more data to settle this problem.

Meanwhile, the pion transition form factor is the most sensitive physical
quantity to observe a non-perturbative contribution to the DA. Other
quantities, for instance, like the pion form factor, may already contain a
hard scattering amplitude at leading order with a soft behavior, due to the
effect of extra coupling constants or gluon propagators \cite{agu}. This means
that they do not lead to such a simple integral over a DA as the one shown by
Eq.(\ref{eqa}). As claimed in Ref.\cite{huang}, we may assume that at present
there is no definite conclusion on which is the asymptotic form of the pion
DA, and it is possible that in the future a combined analysis of data of the
processes involving pions will shed light on the pion distribution amplitude
\cite{huang2}. Notwithstanding, considering the possibility that a flatter
pion DA seems to be favored by the BaBar data
\cite{Radyushkin:2009zg,Polyakov:2009je,Dorokhov:2010zzb,Noguera:2010fe}%
, we can establish a field theoretical limit on how flat this DA can be, and,
consequently, compare this limit to the experimental data. In order to do this
we will study the DA dependence on the non-perturbative dynamics of the
theory, and ultimately on the asymptotic behavior of the pion-quark vertex and
the quark self-energy.

Like in the Nambu-Jona-Lasinio four-fermion approach, in QCD or any
asymptotically free non-Abelian gauge theory, the fermion masses are
dynamically generated along with bound state Goldstone bosons (the pions). The
dynamical quark mass ($\Sigma(p^{2})$), giving by the Schwinger-Dyson equation
is exactly identical to the pseudoscalar Bethe-Salpeter equation (BSE) at zero
momentum transfer ($\Phi_{BS}^{P}(p,q)|_{q\rightarrow0}$), as demonstrated by
Delbourgo and Scadron \cite{ds}
\begin{equation}
\Sigma(p^{2})\approx\Phi_{BS}^{P}(p,q)|_{q\rightarrow0}\,\,\,, \label{eq00}%
\end{equation}
which is a consequence of the fact that they are related through the
Ward-Takahashi identity.

The homogeneous BSE can be, in general, written as
\begin{equation}
\Phi(k,P)= -i\int_{q}^{\infty}\frac{d^{4}q}{(2\pi)^{4}}\, K(k;q,P)\,
S(q_{+})\, \Phi(q;P)\, S(q_{-})\,, \label{eq02}%
\end{equation}
where the amplitude depends on the quarks total ($P$) and relative ($q$)
momenta, with $q_{+}=q+\eta P$, $q_{-}=q-(1-\eta)P$, and $0\leq\eta\leq1$,
where $\eta$ is the momentum fraction parameter. In Eq.(\ref{eq02}) $K$ is the
fully amputated quark-antiquark scattering kernel, $S(q_{i})$ are the dressed
quark propagators, and the homogeneous BSE is valid on-shell, i.e. $P^{2}=0$
in the pion case. Note that we suppressed all indices (color, etc...) in
Eq.(\ref{eq02}).

The BSE, Eq.(\ref{eq02}), is an integral equation that can be transformed into
a second order differential equation. The two solutions of the differential
equation can be found, for example, in Ref.\cite{lane,lan} and are
characterized by one soft asymptotic solution
\begin{equation}
\Phi_{\pi}^{R}(p^{2})\sim\Sigma^{R}(p^{2}>>\mu^{2})\sim\frac{\mu^{3}}{p^{2}%
}\,, \label{eq03}%
\end{equation}
and by the extreme hard high energy asymptotic behavior of a bound state wave
function
\begin{equation}
\Phi_{\pi}^{I}(p^{2})\sim\Sigma^{I}(p^{2}>>\mu^{2})\sim\mu\left[
1+bg^{2}\left(  \mu^{2}\right)  \ln\left(  p^{2}/\mu^{2}\right)  \right]
^{-\gamma}\,, \label{eq1a}%
\end{equation}
where $b=(11N_{c}-2n_{f})/48\pi^{2}$, $c=4/3$ is the Casimir eigenvalue for
quarks in the fundamental representation ($N_{c}=3$ is the number of colors,
and $n_{f}$ is the number of quark flavors), and $\gamma=3c/16\pi^{2}b$. The
asymptotic expression shown in Eq.(\ref{eq1a}) was determined in the appendix
of Ref.\cite{cs} and it satisfies the Callan-Symanzik equation. This last
solution is constrained by the BSE normalization condition \cite{man}, which
imply $\gamma>1/2$, or $n_{f}>5$ \cite{lane,us3}, otherwise it is not
consistent with a possible bound state solution in a $SU(3)$ non-Abelian gauge
theory. We will take $n_{f}=6$ as will be explained later. This solution is one alternative to the soft one
($\Sigma(p^{2})\sim1/p^{2}$) \cite{poli} which leads to the standard DA
$\varphi_{\pi}^{as}(x)$. Nowadays it is known that we may have solutions with
a momentum behavior varying between Eq.(\ref{eq03}) and Eq.(\ref{eq1a})
depending on the theory dynamics \cite{takeuchi,us3}. Note that the Bethe-Salpeter equation (BSE) can be transformed into a second order differential equation. This equation has two possible solutions, one that asymptotically behaves as $1/p^2$ as in Eq.(5) 
and the other one as $[ln (p^2)]^{-\gamma}$ (Eq.(6)). However this result comes out from the homogeneous BSE. The non-homogeneous BSE also includes a normalization condition, as discussed in 
Refs.[27,30,31], that is obeyed by Eq.(5) but when applied to Eq.(6) implies $N_f >5$. This constraint appears because the wave function
is very ``hard", i.e. decreases very slowly with the momentum and cannot be normalized (square integrable) if $\gamma <1/2$. This
condition on $\gamma$ gives the bound $N_f >5$. This limit on $\gamma$ was obtained by Mandelstam in Ref.[30], in QCD for the first
time in Ref.[27] and recently, in a different context, in Ref.[31]. If $N_f<6$ only the solution of Eq.(5) exists, because it would be
the only one obeying the BSE normalization condition. This also means that if $N_f\geq 6$ QCD may have a chiral broken phase whose self-energy is given by Eq.(6). Nowadays it is known that the chiral phase diagram for a non-Abelian theory may change considerably as we change
the number of flavors. For instance, if the theory contains contributions of higher order operators it may have its quark self-energy or bound
state solution varying between Eq.(5) and (6) as discussed in Ref.[33]. We are just saying that if $N_f >5$ Eq.(6) is a possible
bound state solution, and the hardest one that we may have.

It has been argued that Eq.(\ref{eq1a}) may be a realistic wave function in a
scenario where the chiral symmetry breaking is associated to confinement and
the gluons have a dynamically generated mass \cite{us1,us2,us3}. This solution
also appears when using an improved renormalization group approach in QCD,
associated to a finite quark condensate \cite{chan}, and it minimizes the
vacuum energy as long as $n_{f}>5$ \cite{mon}. Moreover, this specific
solution is the only one consistent with Regge-pole like solutions \cite{lan}.
The important fact is that this is the hardest (in momentum space) asymptotic
behavior allowed for a bound state solution in a non-Abelian gauge theory, and
it is exactly for this reason that the constraint on $\gamma$ arises from the
BSE normalization condition. No matter this solution is realized in Nature or
not, it will lead to the flattest pion DA, any other flatter distribution than
this one cannot be a realistic BSE wave function, and would not be consistent
with a composite pion. A totally flat DA can only be related to a fundamental
pion. A realistic DA, in principle, should be related to a solution of the BSE
and should obey a normalization condition peculiar to a well behaved wave function.

The infrared behavior of the gap equation (or BSE) is approximately constant
at small momenta, $\Sigma(p^{2}\rightarrow0)\sim\mu$, where $\mu$, of order of
a few hundred MeV, is the characteristic scale of dynamical quark mass
generation. In order to compute the pion DA we will perform an integral over the wave function in the full
range of momenta (i.e. up to $p^2 \rightarrow \infty $, this is why we will consider $n_f = 6$). To obtain the extreme field theoretical limit on the
pion DA, we shall also work with a simple interpolating expression that roughly
reflects the full behavior of the \textquotedblleft hardest" quark self-energy
(or BSE solution) discussed in the previous paragraphs, namely \cite{us2,us3}
\begin{equation}
\Sigma(p^{2})=\mu\left[  1+bg^{2}\left(  \mu^{2}\right)  \ln\left(
\frac{p^{2}+\mu^{2}}{\mu^{2}}\right)  \right]  ^{-\gamma}\,. \label{eq2}%
\end{equation}
Note that the $\mu$ factor introduced into the logarithm denominator leads to
the right infrared (IR) behavior ($\Sigma(p^{2}\rightarrow0)=\mu$).
Furthermore, this is just one possible ansatz for the full behavior of the self-energy
and other possible interpolations between the IR and ultraviolet (UV)
behaviors are possible, but as long as $\Sigma(p^{2})$ shows the logarithmic
UV behavior our final result will not change. The coupling constant $g^{2}$ is
calculated at the chiral symmetry breaking scale $\mu$, and given by
\begin{equation}
{g}^{2}(k^{2})=\frac{1}{b\ln[(k^{2}+4m_{g}^{2})/\Lambda_{QCD}^{2}]}\,,
\label{eq6}%
\end{equation}
which is an infrared finite coupling determined in QCD where gluons have an
effective dynamical mass $m_{g}$ \cite{cornwall} and is consistent with the
models of Ref.\cite{us3,us1,us2}. $\Lambda_{QCD}$ is the QCD characteristic scale.

Within this approach, the pion distribution amplitude at
leading twist, as a function of the quark self-energy and the pion-quark
vertex, is given by \cite{dor}
\begin{align}
\varphi_{\pi}(x)  &  =\frac{N_{c}}{4\pi^{2}f_{\pi}^{2}}\int_{-\infty}^{\infty
}\frac{d\lambda}{2\pi}\int_{0}^{\infty}du\frac{F(u+i\lambda\bar{x},u-i\lambda
x)}{D(u-i\lambda x)D(u+i\lambda\bar{x})}\nonumber\\
&  \times\left[  x\Sigma(u+i\lambda\bar{x})+\bar{x}\Sigma(u-i\lambda
x)\right]  \,, \label{eq1}%
\end{align}
where the $u$-variable plays the role of the quark transverse momentum
squared, $\lambda x$ and $-\lambda\bar{x}$ are the longitudinal projections of
the quark momentum on the light cone directions ($\bar{x}=(1-x)$), $\Sigma(u)$
is the dynamical quark mass given by (\ref{eq2}),%
\begin{equation}
D(u)\equiv u+\Sigma^{2}(u), \label{Propagator}%
\end{equation}
and the function $F$ is the momentum dependent part of the quark-pion vertex,
which can be approximated by $F(p^{2},{p^{\prime}}^{2})=\sqrt{\Sigma
(p^{2})\Sigma({p^{\prime}}^{2})}$, where $p$ and ${p^{\prime}}$ are the quark
and anti-quark momenta. The pion DA at leading twist is normalized as
\begin{equation}
\int_{0}^{1}dx\,\varphi_{\pi}(x)=1\,. \label{eq1b}%
\end{equation}

It is also useful to write down the expression for the pion DA in the form
found in \cite{Dorokhov:2010zzb} (the so called Schwinger
representation)
\br
\varphi_{\pi}\left(  x\right) &=&\frac{N_{c}}{4\pi^{2}f_{\pi}^{2}}\int%
_{0}^{\infty}\frac{dL}{L}e^{x\overline{x}Lp^{2}}\left[  x\sigma_{m}\left(
xL\right)  \sigma\left(  \overline{x}L\right) \right. \nonumber \\
&+&\left. \left(  x\leftrightarrow
\overline{x}\right)  \right]  , \label{eq1L}%
\er
where $\sigma\left(  \alpha\right)  $ and $\sigma_{m}\left(  \alpha\right)  $
are the Laplace transformations of
\[
\frac{\Sigma^{1/2}(u)}{D(u)}\quad\mathrm{and}\quad\frac{\Sigma^{3/2}(u)}%
{D(u)},
\]
correspondingly. For example,%
\begin{equation}
\frac{\Sigma^{1/2}(u)}{D(u)}=\int_{0}^{\infty}d\alpha e^{-\alpha u}%
\sigma\left(  \alpha\right)  . \label{LT}%
\end{equation}

For the model calculations we take the following parameters: $\mu=100$ MeV,
$\Lambda_{QCD}=300$ MeV and $m_{g}=321.18$ MeV \cite{nat}. To describe Fig. (1) we used $\Lambda_{QCD} = 300$ MeV and for $n_f = 6$ a more appropriate
value would be $\Lambda_{QCD}\approx 200 $ MeV \cite{bethke}. However it should be noticed that the result is more dependent on the ratio
$m_g/\Lambda_{QCD}$ than on the proper $\Lambda_{QCD}$ value. We also emphasize that 
the largest origin of uncertainty in our result is the assumption of Eq. (7) for the self-energy in the full range of momenta. In the intermediate
and infrared region of momenta Eq.(7) may give a poor description of the self-energy, although the good point is that the flat DA behavior is
totally credited to the hard asymptotic self-energy behavior. Within the model
considered we take the expression for the pion decay constant in the so-called
Pagels-Stokar form \cite{Pagels:1979hd}
\begin{equation}
f_{\pi}^{2}=\frac{N_{c}}{4\pi^{2}}\int_{0}^{\infty}du\,\frac{u\Sigma(u)}%
{D^{2}\left(  u\right)  }\left(  \Sigma\left(  u\right)  -\frac{1}{2}%
u\Sigma^{\prime}\left(  u\right)  \right)  , \label{FpiPS}%
\end{equation}
where $\Sigma^{\prime}\left(  u\right)  =d\Sigma\left(  u\right)
/du$. With the given set of parameters we obtain
$f_{\pi}=92.4$ MeV. Actually, it is possible to obtain this $f_{\pi}$ value
with different values for $\mu$ and $m_{g}$, just changing the formula that
interpolates between the IR and UV regimes, although these values should stay
around a few hundreds of MeV.

\begin{figure}[ptb]
\begin{center}
\includegraphics[width=0.9\columnwidth]{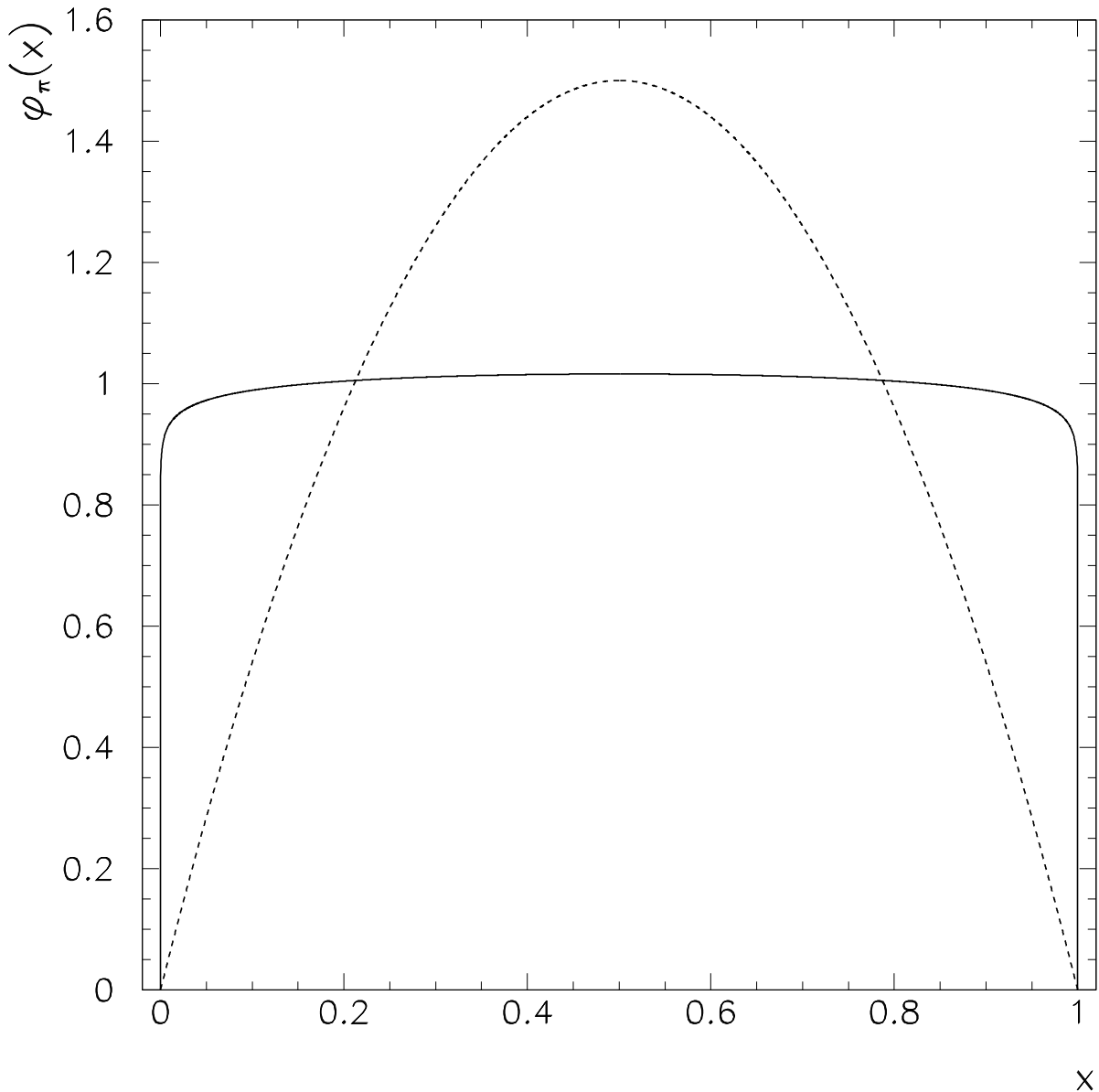}
\end{center}
\caption{Pion DA computed with Eqs.(\ref{eq1}) and (\ref{eq2}), and using the
parameters $m_{g}=322$ MeV and $\mu=100$ MeV (solid curve). The
perturbative-QCD asymptotic pion DA, $\varphi_{\pi}^{as}(x)=6x(1-x)$, is shown
by the dashed curve for comparison.}%
\label{fig1}%
\end{figure}

The pion DA obtained with above parameters and Eqs.(\ref{eq2}) and (\ref{eq1})
is shown in Fig.(\ref{fig1}); for comparison we also draw the asymptotic wave
function $\varphi_{\pi}^{as}(x)$. The DA turns out to be quite flat, and we
have not observed any significant variation as we change $m_{g}$ and $\mu$ as
long as we do not modify the $f_{\pi}$ value in Eq.(\ref{eq1}) and maintain
the logarithmic UV behavior, \textit{which is at the origin of the flat DA behavior}. 
The DA flatness is totally dependent on the logarithmic behavior of the self-energy. 

The asymptotic behavior as $x\rightarrow0$ for the model considered here is
given by %
\begin{equation}
\varphi_{\pi}(x\rightarrow0)\sim\left(  \ln\frac{1}{x}\right)  ^{-\gamma/2} ,
\label{PhiTo0as}%
\end{equation}
where $\varphi_{\pi}(x=0)=\varphi_{\pi}(x=1)=0$. This behavior is actually the
expected one for a pion DA with a vertex function $F(p^{2},p^{\prime2})$
similar to ours, where $F(p^{2},p^{\prime2})$ goes to zero in the limit
$p^{\prime}\rightarrow\infty$ \cite{Dorokhov:2010zzb}. In the 
appendix we determine the asymptotic behavior shown in Eq.(\ref{PhiTo0as}).

Our pion DA numerical result can be reasonably reproduced by using the
normalized form
\begin{equation}
\varphi_{\pi}(x)=\frac{\Gamma\left(  2+2\epsilon\right)  }{\Gamma^{2}\left(
1+\epsilon\right)  }\,x^{\epsilon}(1-x)^{\epsilon}\,, \label{eq3A}%
\end{equation}
where $\epsilon\approx0.024802$. However, it is worth noting that the
calculations performed in this work have been carried out using numerical
values of the pion DA obtained from Eq. (\ref{eq1}).

\begin{figure}[ptb]
\begin{center}
\includegraphics[width=0.9\columnwidth]{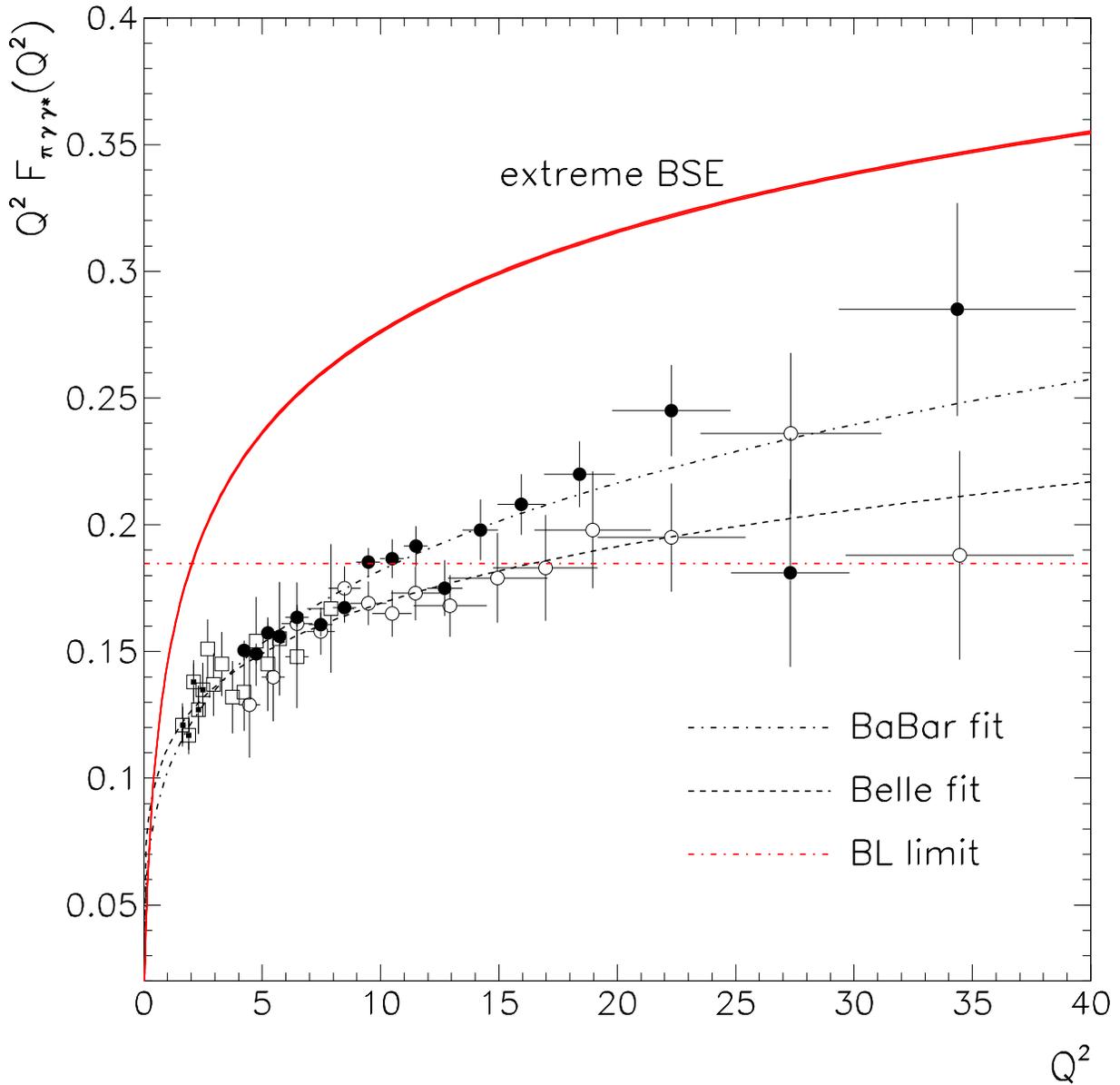}
\end{center}
\caption{The photon-pion transition form factor $F_{\pi\gamma\gamma^{\ast}%
}(Q^{2})$ computed with the extreme DA calculated in this work. The continuous red curve is
the limit obtained with Eq.(\ref{FAsMod}). The horizontal
hatched area correspond to the asymptotic BL limit. The experimental data are taken from CLEO
($\Box$), Belle ($\circ$) and BaBar ($\bullet$) Collaborations.}%
\label{fig2}%
\end{figure}

The leading asymptotic behavior of the form factor is expressed through the
pion DA (\ref{eq3A}) as
\cite{Dorokhov:2010zzb}
\begin{equation}
F_{\pi\gamma^{\ast}\gamma}\left(  0;Q^{2},0\right)  \overset{Q^{2}%
\rightarrow\infty}{=}\frac{2}{3}f_{\pi}\int_{0}^{1}dx\frac{\varphi_{\pi
}\left(  x\right)  }{xQ^{2}}. \label{AsF}%
\end{equation}
If we were considering a totally flat DA this integral would diverge. However,
as emphasized by Radyushkin \cite{Radyushkin:2009zg}, the finite size $R \approx 1/M$ of the
pion should provide a cut-off for the $x$ integral. Therefore the $xQ^{2}$ in the
denominator of Eq.(\ref{AsF}) will be changed as
\be
xQ^{2}\rightarrow xQ^{2} + M^2 \, .
\label{eqxx}
\ee
In principle the factor $M$ should be related to the dynamical quark mass. It was also proposed by
Radyushkin that $M$ could be treated as an effective gluon mass. Indeed the meson radius may have a deep connection with the
effective gluon mass as discussed in \cite{halnat}.
Therefore, no matter we have one case or another, the transition form factor will be giving by
\be
  F_{\pi\gamma^{\ast}\gamma}\left(  0;Q^{2},0\right)  =\frac{2}{3}f_{\pi}\int_{0}^{1}dx\frac{1}{xQ^{2}+ M^2} \, .
\label{FAsMod}%
\ee
$M$ being a dynamical mass should have a momentum dependence showing the decrease of the mass with the momentum.
However when  $xQ^{2}$ is small we can safely substitute
$M(xQ^2)$ by $M$ in Eq.(\ref{FAsMod}), and for large $xQ^{2}$ the momentum behavior of $M(xQ^2)$
is negligible. The result for the transition form factor given by Eq.(\ref{FAsMod}) is shown in Fig.(\ref{fig2}).
In Fig.(\ref{fig2}) we are just assuming $M=320$ MeV no matter this is a quark or gluon dynamical mass.
We are neglecting the effects of the QCD evolution in Eq.(19) and, consequently, in
Fig.(2), because, as verified by Radyushkin \cite{Radyushkin:2009zg}, these effects are small for such a flat DA.

The result of Fig.(\ref{fig2}) for the photon-pion transition form factor can
be compared to the parameterization fit of BaBar \cite{:2009mc} and Belle
\cite{Uehara:2012ag} Collaborations,
\begin{equation}
Q^{2}|F_{\pi\gamma\gamma^{\ast}}(Q^{2})|=A\left(  \frac{Q^{2}}{10~\mathrm{GeV}%
^{2}}\right)  ^{\beta}, \label{eq4}%
\end{equation}
where $A=0.182\pm0.002$ GeV and $\beta=0.25\pm0.02$ for BaBar, and
$A=0.169\pm0.006$ GeV and $\beta=0.18\pm0.05$ for Belle.

With our numerical pion DA we can also compute the second moment of the DA,
which is given by
\begin{equation}
\left\langle \xi^{2}\right\rangle \left(  Q^{2}\right)  =\int_{0}%
^{1}\,dx\,\varphi_{\pi}(x,Q^{2})\xi^{2}\,\,, \label{eq3a}%
\end{equation}
where $\xi=(2x-1)$. Our result can be compared to the lattice value
$\left\langle \xi^{2}\right\rangle =0.28(1)(2)$ \cite{latmoment}, where the
first error is statistical and the second is systematic. The $\left\langle
\xi^{2}\right\rangle $ lattice value has been computed at the scale
$Q^{2}=2.5$ GeV$^{2}$, while our DA calculation is obtained at the scale of
the chiral symmetry breaking. Since the DA evolution with the momentum can
only diminishes the IR value of the integral in Eq.(\ref{eq3a}) we can
consider our result as an upper limit on the value of the second moment of the
DA, i.e.
\begin{equation}
\left\langle \xi^{2}\right\rangle <0.329\,, \label{eq8}%
\end{equation}
which is the highest value for all the parameters that we have considered.

Radyushkin \cite{Radyushkin:2009zg} and Polyakov \cite{Polyakov:2009je} have
proposed a flat behavior for the pion DA in the light of BaBar data. This behavior,
if it indeed happens in Nature, should be justified field theoretically. We have
verified that a flat behavior for the pion DA is totally dependent
on the asymptotic UV logarithmic behavior of the self-energy. When the pion DA is calculated 
with Eq.(\ref{eq1}) using softer quark self-energies the result turns out to be more peaked at $x=0.5$.
The effect of the self-energy behavior at small and intermediate momenta are erased
by the normalization condition Eq.(\ref{eq1b}). 

The high energy limit of the pion transition form factor with a flat DA must have a natural
cut-off ($M$) as proposed by Radyushkin \cite{Radyushkin:2009zg} and Polyakov \cite{Polyakov:2009je},
and this one was introduced by us into Eq.(\ref{FAsMod}). We have assumed a cut-off of the order
of the dynamical masses of the theory, while in Refs.\cite{Radyushkin:2009zg,Polyakov:2009je} these
values are quite high and difficult to associate to some physical scale. This is why our calculation
can only be considered as an upper limit on the pion transition form factor. If we had assumed a
larger $M$ value we would have a better agreement with the BaBar data. Of course, radiative
corrections may also bring down the red curve in Fig.(\ref{fig2}) implying a better adjust of the
data. Hence our result provides a field theoretical limit on the pion DA and
transition form factor based on the hardest asymptotic quark self-energy allowed in QCD. We also
obtained a limit on the second moment of the pion DA compatible with the one of recent QCD lattice simulation.
 
\section*{Acknowledgments}

We are grateful to A. E. Dorokhov for many discussions and the collaboration
in the appendix calculation. This research was partially supported by the Conselho Nacional de
Desenvolvimento Cient\'{\i}fico e Tecnol\'{o}gico (CNPq), by
Coordena\c{c}\~{a}o de Aperfei\c{c}oamento do Pessoal de Ensino Superior
(CAPES), by grant 2013/22079-8 of Funda\c{c}\~{a}o de Amparo \`{a} Pesquisa do
Estado de S\~ao Paulo (FAPESP), and by Funda\c{c}\~{a}o de Amparo \`{a}
Pesquisa do Estado do Rio Grande do Sul (FAPERGS).

\section*{Appendix}

Let us find the small $x$ behavior of $\varphi_{\pi}(x)$ by using the
Schwinger representation (\ref{eq1L}). We also make the following
approximation for the denominators of the integrand%
\[
D\left(  p^{2}\right)  =p^{2}+\Sigma^{2}\left(  p^{2}\right)  \rightarrow
D_{1}\left(  p^{2}\right)  =p^{2}+\mu^{2}.
\]
Then, first we need to find the Laplace transformation of the factor
$\Sigma^{1/2}\left(  p^{2}\right)  /D_{1}\left(  p^{2}\right)  $%
\begin{align*}
&  \frac{\Sigma^{1/2}\left(  p^{2}\right)  }{D_{1}\left(  p^{2}\right)
}=\frac{\mu^{1/2}}{p^{2}+\mu^{2}}\frac{1}{\left[  1+\overline{b}\ln\left(
\frac{p^{2}+\mu^{2}}{\mu^{2}}\right)  \right]  ^{\gamma/2}}\\
&  =\frac{\mu^{1/2}}{p^{2}+\mu^{2}}\frac{1}{\Gamma_{\gamma/2}}\int
d\alpha\alpha^{\gamma/2-1} \times \\
& \,\,\,\,\,\, \exp\left[  -\alpha\left(  1+\overline{b}\ln\left(
\frac{p^{2}+\mu^{2}}{\mu^{2}}\right)  \right)  \right] \\
&  =\frac{\mu^{1/2}}{p^{2}+\mu^{2}}\frac{1}{\Gamma_{\gamma/2}}\int
d\alpha\alpha^{\gamma/2-1}e^{-\alpha}\left(  \frac{\mu^{2}}{p^{2}+\mu^{2}%
}\right)  ^{\alpha\overline{b}}\\
&  =\frac{\mu^{1/2}}{\Gamma_{\gamma/2}}\int d\alpha\alpha^{\gamma
/2-1}e^{-\alpha}\left(  \mu^{2}\right)  ^{\alpha\overline{b}} \times \\
& \,\,\,\,\,\, \frac{1}%
{\Gamma\left(  1+\alpha\overline{b}\right)  }\int d\beta e^{-\beta\left(
p^{2}+\mu^{2}\right)  }\beta^{\alpha\overline{b}}\\
&  =\frac{\mu^{1/2}}{\overline{b}^{\gamma/2}}\int d\beta e^{-\beta p^{2}%
}e^{-\beta\mu^{2}} \times \\
& \,\,\,\,\,\,\, \int_{0}^{\infty}d\alpha e^{-\alpha/\overline{b}}%
\frac{\alpha^{\gamma/2-1}\left(  \mu^{2}\beta\right)  ^{\alpha}}%
{\Gamma_{\gamma/2}\Gamma\left(  1+\alpha\right)  }\equiv\int d\beta e^{-\beta
p^{2}}G_{1/2}\left(  \beta\right)  ,
\end{align*}
where $\bar{b}\equiv bg^{2}$ and%
\[
G_{1/2}\left(  \beta\right)  =e^{-\beta\mu^{2}}\frac{\mu^{1/2}}{\overline
{b}^{\gamma/2}}\frac{1}{\Gamma_{\gamma/2}}\int_{0}^{\infty}d\alpha
e^{-\alpha/\overline{b}}\frac{\alpha^{\gamma/2-1}\left(  \mu^{2}\beta\right)
^{\alpha}}{\Gamma\left(  1+\alpha\right)  }.
\]
Similarly we obtain the Laplace transform of the factor
$\Sigma^{3/2}\left(  p^{2}\right)  /D_{1}\left(  p^{2}\right)  $ as%
\[
G_{3/2}\left(  \beta\right)  =e^{-\beta\mu^{2}}\frac{\mu^{3/2}}{\overline
{b}^{3\gamma/2}}\frac{1}{\Gamma_{3\gamma/2}}\int_{0}^{\infty}d\alpha
e^{-\alpha/\overline{b}}\frac{\alpha^{3\gamma/2-1}\left(  \mu^{2}\beta\right)
^{\alpha}}{\Gamma\left(  1+\alpha\right)  }.
\]
Next we substitute these expressions in (\ref{eq1L}) keeping in mind that we
are interested in the $x\rightarrow0$ behavior. Then one has for the integrand%
\begin{align*}
&  \overline{x}G_{0,m}\left(  xL,\overline{x}L\right)  =\overline{x}%
G_{1/2}\left(  xL\right)  G_{3/2}\left(  \overline{x}L\right) \\
&  \overset{x\rightarrow0}{\approx}e^{-L\mu^{2}}\frac{\mu^{2}}{\overline
{b}^{2\gamma}\Gamma_{\gamma/2}\Gamma_{3\gamma/2}} \times \\
& \,\,\,\,\,\, \int_{0}^{\infty}d\alpha
e^{-\alpha/\overline{b}}\frac{\alpha^{\gamma/2-1}\left(  \mu^{2}xL\right)
^{\alpha}}{\Gamma\left(  1+\alpha\right)  } \times \\
& \,\,\,\,\,\, \int_{0}^{\infty}d\beta
e^{-\beta/\overline{b}}\frac{\beta^{3\gamma/2-1}\left(  \mu^{2}L\right)
^{\beta}}{\Gamma\left(  1+\beta\right)  }%
\end{align*}
and for the pion DA%
\begin{align*}
&  \varphi_{\pi}\left(  x\rightarrow0\right)  =\frac{N_{c}}{4\pi^{2}f_{\pi
}^{2}}\int_{0}^{\infty}\frac{dL}{L}G_{0,m}\left(  xL,\overline{x}L\right) \\
&  =\frac{N_{c}}{4\pi^{2}f_{\pi}^{2}}\int_{0}^{\infty}\frac{dL}{L}e^{-L\mu
^{2}}\frac{\mu^{2}}{\overline{b}^{2\gamma}\Gamma_{\gamma/2}\Gamma_{3\gamma/2}%
} \times \\
& \,\,\,\,\,\, \int_{0}^{\infty}d\alpha e^{-\alpha/\overline{b}}\frac{\alpha^{\gamma
/2-1}\left(  \mu^{2}xL\right)  ^{\alpha}}{\Gamma\left(  1+\alpha\right)  } \times \\ %
& \,\,\,\,\,\, \int_{0}^{\infty}d\beta e^{-\beta/\overline{b}}\frac{\beta^{3\gamma
/2-1}\left(  \mu^{2}L\right)  ^{\beta}}{\Gamma\left(  1+\beta\right)  }\\
&  =\frac{N_{c}}{4\pi^{2}f_{\pi}^{2}}\frac{\mu^{2}}{\overline{b}^{2\gamma
}\Gamma_{\gamma/2}\Gamma_{3\gamma/2}}\int_{0}^{\infty}d\alpha e^{-\alpha
/\overline{b}}\alpha^{\gamma/2-1}x^{\alpha}f\left(  \alpha\right)  ,
\end{align*}
where
\[
f\left(  \alpha\right)  =\int_{0}^{\infty}d\beta e^{-\beta/\overline{b}}%
\beta^{3\gamma/2-1}\frac{\Gamma\left(  \alpha+\beta\right)  }{\Gamma\left(
1+\alpha\right)  \Gamma\left(  1+\beta\right)  }.
\]

In the asymptotic regime $x\rightarrow0$, we can take $f\left(  \alpha\right)
$ at $\alpha=0$ and obtain
\begin{equation}
\varphi_{\pi}\left(  x\rightarrow0\right)  \sim\frac{N_{c}}{4\pi^{2}f_{\pi
}^{2}}\frac{\mu^{2}}{\overline{b}\left(  3\gamma/2-1\right)  }\left(
1+\overline{b}\ln\frac{1}{x}\right)  ^{-\gamma/2}. \label{PhiAs} \nonumber
\end{equation}
This is the result given above in equation (\ref{PhiTo0as}). Note, that this
last result is proportional to $\Sigma^{1/2}$ with the argument
$\left(  p^{2}+\mu^{2}\right)  /\mu^{2}$ of log substituted by $1/x$. If
$\gamma\leq2$ then the convergence to this asymptotic behavior is rather slow.

\end{document}